\documentclass[5p,twocolumn]{elsarticle}

\journal{Physics Letters B}

\usepackage{leftindex}
\usepackage{graphicx}
\usepackage{amsfonts}
\usepackage{amsmath,amssymb,tensor,bm,bbm}
\usepackage[usenames]{color}
\usepackage{graphics}
\usepackage{marginnote}
\usepackage{latexsym}
\usepackage{rotating}
\usepackage{float}
\usepackage{hyperref}
\usepackage{xspace} 
\usepackage[utf8]{inputenc} 
\usepackage[T1]{fontenc}    
\usepackage{lmodern}        

\usepackage{slashed}

\allowdisplaybreaks

\def\p{\partial}

\def\bi{\begin{itemize}}
\def\ei{\end{itemize}}
\def\be{\begin{equation}}
\def\ee{\end{equation}}
\newcommand{\bea}{\begin{eqnarray}}
\newcommand{\eea}{\end{eqnarray}}

\begin{document}
\begin{frontmatter}
\title{Birefringence in fermion-attenuated gravitational wave power spectrum}

\author[Fudan,FudanAstro]{Jinglong Liu}

\author[Brown]{Stephon Alexander}

\author[Fudan,FudanAstro,INFN1,INFN2]{Antonino Marcian\`o}

\address[Fudan]{Center for Field Theory and Particle Physics \& Department of Physics, Fudan University, 200433 Shanghai, China}
\address[FudanAstro]{Center for Astronomy and Astrophysics, Fudan University, 200433 Shanghai, China}

\address[Brown]{Brown Theoretical Physics Center and Department of Physics,
Brown University,
RI 02903, USA}

\address[INFN1]{Laboratori Nazionali di Frascati INFN, Frascati (Rome), Italy, EU}
\address[INFN2]{INFN sezione Roma Tor Vergata, I-00133 Rome, Italy, EU}

\begin{abstract}
\noindent 
Within the framework of Chern-Simons gravity, a theory that dynamically violates parity, we analyze the power spectrum of gravitational waves in light of the damping effect due to the free streaming relativistic neutrinos and dark fermions. The power spectrum is expressed terms of right- and left-handed polarizations, and the evolution of the gravitational waves is studied numerically. Birefringence is explicitly shown in the power spectrum, though the difference in the amplitudes is small. Specific features of peaks and dips appear gravitational wave power spectrum mirroring chiral gravitational wave mediated parametric resonance during reheating. Our result represents a useful tool to test Chern-Simons gravity and enables to constrain mechanisms of inflation and reheating related to this theory. We predict a falsifiable pattern of observable peaks and dips in the chiral independent  gravitational power spectrum, eventually observable in next space-borne gravitational interferometers, including LISA, Taiji and Tianqin.

\end{abstract}

\end{frontmatter}

\section{Introduction}
\noindent
Gravitational waves (GWs), predicted by Einstein as a feature of General Relativity and directly observed almost a decade ago \cite{LIGOScientific:2016aoc}, have become a new powerful probe for both the early and late universe. Direct detection of GWs from binary systems achieved through terrestrial interferometers \cite{LIGOScientific:2016aoc}, and indirect measurements obtained by radio telescopes observing pulsar timing arrays \cite{Ballmer:2022uxx,Kalogera:2021bya,NANOGrav:2023hvm}, have opened a new observational channel to test departure from General Relativity and particle theories beyond the standard model. Particularly relevant for the  phenomenology of high energy particle physics, extended theories of gravity and dark matter models are stochastic gravitational waves (SGWs) \cite{Christensen:2018iqi}. Their sources include the inflationary epoch, phase transitions in the early universe, alternative cosmologies, and cosmic strings. They can also be produced at astrophysical scales, from supernovae, magnetars, neutron stars or black holes, over the history of the universe. Thus SGWs provide a good candidate to understand the physics of the early universe. 

Once GWs are produced, they propagate almost freely. However, they still interact with particles, such as relativistic axions \cite{Dent:2013asa,Ringwald:2020vei}, nonrelativistic and collisional matter \cite{Baym:2017xvh,Flauger:2017ged,Miron-Granese:2020hyq,Zarei:2021dpb}, and vector fields \cite{Bielefeld:2015daa,Miravet:2020kuj,Tishue:2021blv,Miravet:2022pli}. One interesting effect is provided by free-streaming neutrinos, as proposed by Steven Weinberg in Ref.~\cite{Weinberg:2003ur}. In general, the relativistic fermions with self-interactions will take part in the damping effect for either the inflationary GWs re-entering the horizon or causal GWs \cite{Hook:2020phx,Brzeminski:2022haa} --- see e.g. the low-frequency tail of stochastic backgrounds generated by subhorizon processes, such as phase transitions or resonant particle productions.

In recent years, the possibility that parity violation may arise in the gravitational sector has been attracting increasing attention. Chern-Simons (CS) gravity \cite{jackiw2003chern,ALEXANDER20091} represents a paradigmatic framework to inspect this scenario. Within this framework \cite{Lue:1998mq,Alexander:2004us,Alexander:2014bsa}, GWs are expected to show a peculiar feature of birefringence, with right-handed gravitational waves amplified while left-handed turning out to be damped. In Ref.~\cite{Alexander:2022cow}, an effective theory with self-interacting fermions and gravitons has been found to reproduce CS gravity\footnote{An effective theory of self-interacting fermions was extensively studied in the literature \cite{1971PhLA...36..225H,hehl1973spin,hehl1974spin,Hehl:1976kj,lord1976tensors,doi:10.1142/0233,de1994spin,Shapiro:2001rz,hammond2002torsion}, and can be obtained by integrating out the torsional component of the Lorentz connection in the Einstein-Cartan-Holst action coupled to Dirac fermions. The related cosmological and astrophysical implications were investigated in Refs.~\cite{Alexander:2014eva,Bambi:2014uua,Alexander:2014uaa,Addazi:2016rnz,Addazi:2017qus,Addazi:2018zjv}. }. It is tempting to investigate in this context the damping effect of GWs as sourced by parity violation in the gravitational sector. We combine the analysis of the damping effect on GWs, as sourced by the interaction with relativistic fermions, with the phenomenological investigation of CS gravity. In particular, we derive the power spectrum of GWs of left and right circularly polarization states. This observable would indeed represent a significant probe of the early universe, including inflation and reheating in CS gravity. As a matter of fact, parametric resonance mediated by GWs of different helicities were already studied in Ref.~\cite{Alexander:2014bsa} for their role during reheating.

In Sec.~\ref{sec:CS} we review GWs birefringence in CS gravity. In Sec.~\ref{sec:damp} we discuss the damping effect provided by self-interacting relativistic fermions. Numerical result are provided in Sec.~\ref{sec:nu}. Finally, in Sec.~\ref{conclu} we provide a discussion of the results obtained and spell our conclusions.

\section{Dynamical Chern-Simons gravity and gravitational waves birefringence}\label{sec:CS}
\noindent
To describe chirality in GWs, we introduce CS gravity. This is a modified theory of General Relativity in 4 dimension proposed by Jackiw and Pi \cite{jackiw2003chern} ---  for a comprehensive review see e.g. Ref.~\cite{ALEXANDER20091}. CS gravity is defined by the action 
\begin{equation}
    S = S_{\rm EH} + S_{\rm CS} + S_\phi + S_{\rm matter},\label{LCS1}
\end{equation}
where
\begin{equation}
    S_{\rm EH} = \frac{\kappa}{2} \int dx^4 \sqrt{-g}R\label{LCS2}
\end{equation}
is the Einstein-Hilbert action with $\kappa = 1/(8\pi G)$, 
\begin{equation}
    S_{\rm CS} = \frac{\alpha}{4} \int dx^4 \sqrt{-g} \phi (\leftindex^*{R}R)\label{LCS3}
\end{equation}
is the CS deformation with a pseudoscalar $\phi$ and a Pontryagin density $\leftindex^*{R}R = R\tilde{R} = \tensor[^*]{R}{^a_b^c^d}\tensor{R}{^b_a_c_d}$, with $*$ denoting the gravitational Hodge dual.
\begin{equation}
    S_\phi = -\frac{\beta}{2}\int dx^4\sqrt{-g}(g^{ab}\nabla_a\phi\nabla_b\phi+V(\phi))\label{LCS4}
\end{equation}
is the dynamical action for the pseudoscalar $\phi$, and $S_{\rm matter}$ is the action for matter. The dimensions of the coupling constants $\alpha$ and $\beta$ are determined by the dimension, hence the dynamics, of the pseudoscalar $\phi$. We select a dynamical CS gravity that corresponds to the specific choice such that $\beta = 1$ and $\alpha$ have dimension of length, as already discussed in Refs.~\cite{Choi:1999zy,Alexander:2014bsa}. We consider the simplified form of the potential $V(\phi)=\frac{1}{2}m^2\phi^2$, and exploit CS gravity to study the role of the pseudoscalar as inflaton in early cosmology. The action considered is the extended chaotic inflation model with CS gravity discussed in \cite{Alexander:2014bsa}. The equations of motion for the metric tensor and the pseudoscalar are 
\begin{align}
    &G_{\mu\nu} + 16\pi G\alpha C_{\mu\nu} = 8\pi T_{\mu\nu},\\
    &\Box\phi - m_\phi^2\phi +\frac{\alpha}{4}R\tilde{R} = 0,
\end{align}
where $\Box = \p_t^2 + 3H\p_t - \frac{1}{a^2}\nabla^2$, and 
\begin{align}
    C^{\mu\nu} = \nabla_\alpha \epsilon^{\alpha\beta\gamma(\mu}\nabla_\gamma R_\beta^{\nu)}+\nabla_{(\alpha}\nabla_{\beta)}\phi\tilde{R}^{\beta(\mu\nu)\alpha}.
\end{align}
CS gravity introduces at the cosmological level GWs birefringence, as widely discussed in the literature --- see e.g. Refs.~\cite{Lue:1998mq, Alexander:2004us,Alexander:2004wk,Alexander:2014bsa}. Parity violation originates from the gravitational sector and hence percolates in the GWs, which acquire different amplitudes for their left/right-handed components.
To investigate the propagation of GWs, we focus on the Friedmann-Lema\^{i}tre-Robertson-Walker (FLRW) metric in conformal time $\tau$, and its tensor perturbations $h_{ij}$, i.e. 
\begin{align}
    ds^2 = a^2(\tau)\left(-d\tau^2 +[\delta_{ij}+h_{ij}(\tau,\mathbf{x})]dx^idx^j\right)\,. \label{FRWmetric}
\end{align}
Tensor perturbations are assumed to be transverse and traceless (TT), i.e. $\p^ih_{ij}=0$ and $h^i_{\ i}=0$. Their equation of motion has been already dealt with in the literature, while addressing dynamical CS gravity--- see e.g. Ref.~\cite{Choi:1999zy} --- and reads
\begin{align}
    \bar{\square} h^j_{\ i} = &\frac{4\alpha}{a^2M_{\rm Pl}^2}\tilde{\epsilon}^{pk(j}\left[\left(\phi''-2\mathcal{H}\phi'\right)\p_p h_{i)k}'+\phi'\bar{\Box}\p_ph_{i)k}\right] \notag\\
    &+ \frac{2a^2}{M_{\rm Pl}^2}\pi^T_{ij},
\end{align}
where $\pi_{ij}^T$ is the anisotropic stress-energy tensor, representing the transverse and traceless component of the stress-energy tensor perturbation $\delta T_{ij}$, $M_{\rm pl}$ is the reduced Planck mass, $\bar{\square}$ is the D'Alembertian operator in conformal time $\tau$ associated with the background, namely
\begin{align}
    \bar{\square}=\partial_\tau^2 + 2\mathcal{H}\partial_\tau - \delta^{ij}\partial_i\partial_j\,.
\end{align}
To simplify the equation of motion, one can follow Ref.~\cite{Alexander:2014bsa} and use order reduction. The tensor perturbation can be decomposed into $h_{\mu\nu} = h_{ij}^{\rm GR} + \alpha^2 \delta h_{ij}$, i.e. the product of a GR level perturbation and a beyond GR level perturbation, where $h_{ij}^{\rm GR}$ satisfies $R_{\mu\nu}[h^{\rm GR}_{ij}]=0$. Recasting $h_{ij}$ in the helicity basis $\epsilon^{R/L}_{ij} = (\epsilon^+_{ij} \pm i\epsilon_{ij}^\times)/\sqrt{2}$,
\begin{align}
    h_{ij} = \frac{1}{(2\pi)^{3/2}}\int dk^3\sum_s \epsilon^s_{ij}h^s_k(\eta)e^{i\mathbf{k}\cdot \mathbf{x}}\,, \quad s\!=\!L,R\,,
\end{align}
the equation of motion becomes
\begin{align}
    \bar{\square} h_{R/L} = \pm i\frac{4\alpha}{a^2M_{\rm Pl}^2}\left(\phi''-2\mathcal{H}\phi'\right)\p_zh_{R/L}'+\frac{2a^2}{M_{\rm Pl}^2}\pi^T_{R/L},
\end{align}
where on the right-hand side the `+' sign denotes the right-handed component $h_R$, the `-' sign denotes the left-handed component $h_L$, and we have assumed the GW to propagate on the $z$ direction. In momentum space, the aforementioned equation of motion becomes
\begin{align}
    &h''_R + 2\mathcal{H} h'_R + k^2 h_R = -\Theta k h_R' + \frac{2a^2}{M_{\rm pl}^2}\pi_R^T\,,   \label{EOMhR}\\
    &h''_L + 2\mathcal{H} h'_L + k^2 h_L = \Theta k h_L' + \frac{2a^2}{M_{\rm pl}^2}\pi_L^T\,,       \label{EOMhL}
\end{align}
where we have defined 
\begin{align}
    \Theta = \frac{2\alpha}{M_{\rm Pl}^2 a^2}(\phi''-2\mathcal{H}\phi')\,,    \label{CSterm}
\end{align}
prime denoting the derivative with respect to the conformal time, and $\mathcal{H}=a'(\tau)/a(\tau)$ the Hubble parameter in conformal time.

\section{Damping of the gravitational waves}\label{sec:damp}
\noindent 
GWs interact with relativistic particles. After the weak interaction is decoupled, the standard model neutrinos start free streaming, and the GWs' amplitude is reduced by around 20\% \cite{Weinberg:2003ur}. Recently, in Ref.~\cite{Loverde:2022wih}, changes of GWs amplitude and power spectrum due to either neutrinos or other dark fermions provided with self-interactions have been discussed. We exploit similar methods in order to investigate the dumping effect originated by the self-interacting fermions in the anisotropic stress-energy tensor. 

The Boltzmann equation provides a powerful way to evaluate the anisotropic stress-energy tensor for fermions --- see e.g. the recent analysis in Ref.~\cite{Loverde:2022wih}, following up on Ref.~\cite{Ma:1995ey}. The Boltzmann equation reads
\begin{align}
    \frac{d}{d\tau} f(\tau,\mathbf{x},q,\hat{q}) = \mathcal{C}[f]\,,
\end{align}
$f(\tau,\mathbf{x},q,\hat{q})$ denoting the phase-space density, depending on the space coordinate $\boldsymbol{x}$, the conformal time $\tau$, the co-moving momentum and the momentum direction $\hat{q}$, and $\mathcal{C}[f]$ denoting the collision term. The density can be expanded around a background $\bar{f}(\tau,\mathbf{x},q,\hat{q})$, 
\begin{align}
    f(\tau,\mathbf{x},q,\hat{q}) = \bar{f}(\tau,q)[1+\Psi(\tau,\mathbf{x},q,\hat{q})],
\end{align}
$\Psi(\tau,\mathbf{x},q,\hat{q})$ denoting a perturbation. The Boltzmann equation then becomes
\begin{align}
    \frac{\p\ln\bar{f}}{\p\tau} + \frac{\ln\bar{f}}{\p q}\frac{dq}{d\tau} + \frac{d\ln\bar{f}}{d\tau}\bar{\Psi}+\frac{d\Psi}{d\tau} = \frac{1}{\bar{f}}\mathcal{C}[f]\,. 
\end{align}
At the background level $dq/d\tau$ vanishes, and one can assume the $\bar{f}$'s dependence in time to be negligible. These conditions are satisfied by relativistic species at the equilibrium or collisionless particles. As a result, one can ignore the parts corresponding to $d\bar{f}/d\tau$ in the Boltzmann equation. Thus the perturbed equation becomes 
\begin{align}
    \frac{d\Psi}{d\tau} + \frac{\partial \ln\bar{f}}{\partial q}\delta\left(\frac{dq}{d\tau}\right) + \Psi \frac{d\ln\bar{f}}{d\tau} = \frac{1}{\bar{f}}\delta\mathcal{C}[f]\,,
\end{align}
where the $\delta$ terms denote perturbations to the background part. The distribution function $\Psi$ in momentum space can also be decomposed into polarization tensors, namely
\begin{align}
    \Psi = \Psi^{(S)} + \sum_\lambda \epsilon_\lambda^i \hat{q}_i \Psi^{(V)}_\lambda + \sum_\lambda \epsilon^{ij}_\lambda \Psi_\lambda^{ij}\,.
\end{align}
Correspondingly, the Boltzmann equation, proportional to the tensor polarization, is expressed by
\begin{align}\label{Bolper}
    \frac{\p\Psi^T_\lambda}{\p\tau} + ik\mu \frac{q}{E}\Psi^T_\lambda - \frac{1}{2}\frac{\p\ln\bar{f}}{\p\ln q}h'_\lambda = \frac{1}{\bar{f}(q)}\mathcal{C}[f]_\lambda^T\,,
\end{align}
where $\mu = \hat{k}\cdot \hat{q}$ is the inner product between the wave number and the direction of the GW propagation, $E$ is the co-moving energy $E=a\sqrt{p^2 + m^2}$, and for relativistic particles $p\gg m$ and $E\approx q$. The dependence of $\mu$, i.e. the cosine of the angle between $\hat{k}$ and $\hat{q}$, enables to use moment expansion to decompose the functions $\Psi$ and $\mathcal{C}$ onto the orthogonal bases of the Legendre polynomials $P_l(\mu)$. Thus, after suitably normalization, one finds 
\begin{align}
    F(\tau,\mathbf{k},\hat{q})&\equiv \frac{\int dq q^3\bar{f}(q)\Psi(\tau,\mathbf{k},q,\hat{q})}{\int dq q^3\bar{f}(q)}\nonumber\\
    &= \sum_{l=0}^\infty(-i)^l(2l+1)F_l(\tau,\mathbf{k})P_l(\mu)\,,\\
    C(\tau,\mathbf{k},\hat{q})&\equiv \frac{\int dq q^3 \mathcal{C}[f]/\bar{f}(q)}{\int dq q^3\bar{f}(q)}\nonumber\\
    &= \sum_{l=0}^\infty (-i)^l(2l+1)C_l(\tau,\mathbf{k})P_l(\mu).
\end{align}
Using the recurrence relation for the Legendre polynomials $(l+1)P_{l+1}(\mu) = (2l+1)\mu P_l(\mu) - lP_{l-1}(\mu)$, equation \eqref{Bolper} becomes a Boltzmann hierarchy 
\begin{align}
    \frac{\p F_{\lambda,l}^{(T)}}{\p\tau}-\frac{k}{2l+1}\bigl[lF_{\lambda,l-1}^{(T)}-(l+1)F_{\lambda, l+1}^{(T)}\bigr]+2\delta_{l0}h'_\lambda=C_{\lambda,l}^{(T)}.\label{BoltHier}
\end{align}
As for the stress-energy tensor, this can be recast in terms of the phase space distribution function as 
\begin{equation}
    T^\alpha_{\ \beta}(\tau,\mathbf{k})=\frac{1}{\sqrt{-g}}\int dp_1dp_2dp_3\frac{p^\alpha p_\beta}{p^0}f(\tau,\mathbf{k},q,\hat{q})\,.
\end{equation}
Thus the TT component of the space perturbation projected on $\epsilon_{ij}^\lambda$ reads  
\begin{equation}\label{anisoSE}
    \pi^T_\lambda(\tau,\mathbf{k})=\bar{\rho}_\nu(\tau)\biggl(\frac{2}{15}F_{\lambda,0}^{(T)}+\frac{4}{21}F_{\lambda,2}^{(T)}+\frac{2}{35}F_{\lambda,4}^{(T)}\biggr)\,,
\end{equation}
where $\bar{\rho}_\nu(\tau)$ is the neutrinos' background energy density. Performing the substitution $\bar{\rho}_\nu = \Omega_\nu \bar{\rho}$, and then exploiting the Friedmann equation $\mathcal{H}^2(\tau) = a^2(\tau)\bar{\rho}/(3M_{\rm pl}^2)$, the matter part in the equation of motion \eqref{EOMhR} and \eqref{EOMhL} becomes 
\begin{align}
    \frac{2a^2}{M_{\rm pl}^2}\pi_\lambda^T = 6\Omega_\nu(\tau)\mathcal{H}^2\left(\frac{2}{15}F_{\lambda,0}^{(T)}+\frac{4}{21}F_{\lambda,2}^{(T)}+\frac{2}{35}F_{\lambda,4}^{(T)}\right).
\end{align}
The collision term for the 2-2 scattering is provided according to Ref.~\cite{Oldengott:2017fhy}, i.e.
\begin{equation}\label{collisionterm}
    C_{\lambda,l}=\alpha_l\p_\tau\kappa_\nu F_{\lambda,l},
\end{equation}
where $\kappa_\nu$ is the optical depth of the species $\nu$, $\p_\tau\kappa_\nu$ is the time-dependent interaction rate, and $\alpha_l$ are numerical coefficients. In Ref.~\cite{Loverde:2022wih}, the optical depth can be modeled in terms of the temperature, and be provided by the expression
\begin{align}
    \p_\tau \kappa_\nu = \frac{a(\tau)}{a(\tau_\star)}\lambda T_\nu^n(\tau),
\end{align}
where $T_\nu$ is the temperature of the fermions, $n$ labels the interaction strength, and $\lambda$ is an effective coupling constant of mass dimension $1-n$. The footnote $\star$ denotes the de-/recoupling time. The de-/recoupling happens at the time when the interacting rate equals the Hubble parameter, $|\p_\tau \kappa_\nu(\tau_\star)| = \mathcal{H}(\tau_\star)$. Using this condition, one can modulate the optical depth as 
\begin{align}
    \p_\tau\kappa_\nu(\tau) = \frac{\lambda T_\nu^n(\tau_\star)}{[a(\tau)/a(\tau_\star)]^{(1-n)}} = -k_\star\left(\frac{a(\tau)}{a(\tau_\star)}\right)^{1-n}.
\end{align}
Applying the Friedmann equation 
\begin{align}
    3M_{\rm pl}^2\mathcal{H}^2(\tau) = a^2(\tau)\frac{\bar{\rho}_\nu(\tau)}{\Omega_\nu(\tau)},
\end{align}
where $\bar{\rho}_\nu(\tau)$ is the energy density of the fermion species labeled by $\nu$, and $\Omega_\nu(\tau)$ is its energy fraction, the interaction rate can be evaluated in terms of $\tau_k = 1/k$ as 
\begin{align}\label{InterRatek}
    \p_\tau\kappa_\nu(\tau) = -k\left(\frac{a(\tau)}{a(\tau_k)}\right)^{1-n}\left(\frac{k}{k_\star}\right)^{n-2}\left(\frac{\Omega_\nu(\tau_k)}{\Omega_\nu(\tau_\star)}\right)^{(n-1)/2},
\end{align}
or in terms of the phase transition time $\tau_i$ as  
\begin{align}\label{InterRatei}
    \p_\tau \kappa_\nu(\tau) = -k_i\left(\frac{a(\tau)}{a(\tau_i)}\right)^{1-n}\left(\frac{k_i}{k_\star}\right)^{n-2}\left(\frac{\Omega_\nu(\tau_i)}{\Omega_\nu(\tau_\star)}\right)^{\frac{n-1}{2}}.
\end{align}
The aforementioned relations will find use in what follows. 

\section{Numerical results}\label{sec:nu}
\noindent
We are now ready to discuss the GW solutions that in the presence self-interacting fermions encode parity violation. The equations of motion we are solving are \eqref{EOMhR}, \eqref{EOMhL}, and the Boltzmann hierarchy \eqref{BoltHier}, with the collision term provided in Eq.~\eqref{collisionterm}. To discuss numerical solutions of the aforementioned equations, we implement a similar procedure as in Ref.~\cite{Loverde:2022wih}. We may break down the steps adopted to derive our results. We start with the truncation of the Boltzmann hierarchy at $l_{\rm max}=100$, and use
\begin{align}
    F_{l_{\rm max}} = \frac{2l_{\rm max} - 1}{k\tau} F_{l_{\rm max}-1} - F_{l_{\rm max}-2}\,.
\end{align}
To find the solutions, we use the function \texttt{solve\_ivp} in the \textsf{SciPy}'s package \cite{Virtanen:2019joe} and apply it separately for $h_{L/R}$ and $F_{\lambda, l}$ in an iterative way. We solve the differential equations using the Radau IIA routine \cite{wanner1996solving}, with relative tolerance $10^{-6}$ and absolute tolerance $10^{-9}$, for $F_{\lambda l}$, and $10^{-15}$ for $h_{L/R}$ and $\p_x h_{L/R}$. To find iterative solutions, we focus on the equations for $h_{L/R}$, with initial $F_{\lambda,l}=0$, then substitute into the Boltzmann hierarchy to recover non-vanishing $F_{\lambda,l}$, and so on. For all the cases under scrutiny, we find that iterating up to five times provides enough accuracy. To compare GWs amplitude results referring to alternative scenarios, either accounting for fermions or for a CS term or both of them, we need to need to provide the amplitudes' expression. Since the oscillatory damping ends when the fermions stop interacting with the GWs, the anisotropic stress-energy tensor in Eq.~\eqref{EOMhR} vanishes, the CS term being negligible at late cosmological times, being dependent on the time-derivative of the inflaton. As a result, the amplitude of the left- right-handed GWs $h_{L/R}$ can be recovered by matching the solutions to the ones not accounting for an anisotropic stress-energy tensor or a CS term. In the approximation used in Ref.~\cite{Loverde:2022wih}, this entails
\begin{align}
    A^\infty_\lambda \approx a(x_f/k)\sqrt{h_\lambda^2(x_f)+\p_x h_\lambda^2(x_f)}\,,
\end{align}
where $x_f$ is the final time, which is large enough to correspond to modes deep inside the horizon, and the damping in the GWs' power spectrum can be expressed by 
\begin{align}
    \frac{\Omega_{\rm GW}}{\Omega_{\rm GW}[f_{\rm fs}=0]} = \left(\frac{A_\infty}{A_\infty[f_{\rm fs}=0]}\right)^2\,,\label{ChiralS}
\end{align}
where $f_{\rm fs}$ denotes the energy fraction of the free streaming particles under investigation. We claim that Eq.~\eqref{ChiralS} is used to calculate the damping effect of the GWs' power spectrum with any specific chirality, while the total damping spectral abundance can be recovered as the average over both chiralities. To discuss the values of the CS parameter $\Theta$, we exploit the solution of the inflaton 
\begin{align}
    \phi(\tau) \approx \phi_0(\tau) \sin(m_\phi a\tau), \qquad \phi_0(\tau) = \frac{M_{\rm pl}}{m_\phi a \tau},
\end{align}
and substitute it into Eq.~\eqref{CSterm}, hence obtaining 
\begin{align}
    \Theta = &-\frac{2\alpha\phi_0 m_\phi^2}{M_{\rm pl}^2}\sin(m_\phi a \tau)-\frac{4\alpha \phi_0 H m_\phi}{M_{\rm pl}^2}\cos(m_\phi a\tau)\notag\\
    = &-\frac{2\alpha H m_\phi}{M_{\rm pl}}\sin(m_\phi a \tau) - \frac{4\alpha H^2}{M_{\rm pl}}\cos(m_\phi a\tau).
\end{align}
During the Universe radiation dominated era, when propagating modes re-enter the horizon during reheating, the inflaton oscillates fast enough to mediate the resonant particle production. Thus the oscillation rate has to exceed the expansion rate of the Universe, i.e., $m_\phi a \gg \mathcal{H}$, in order to ignore the second term in $\Theta$. While, once the first term is substituted into the equations of motion \eqref{EOMhR} and \eqref{EOMhL}, both the sides of which having been divided by $k^2$, and the dimensionless modes $x=k\tau$ have been defined, we find 
\begin{align}\label{FinalEOMh}
    \p_x^2 h_{R/L} + 2\frac{\p_x a}{a}\p_x h_{R/L} =& \pm\left[\frac{2\alpha H m_\phi}{M_{\rm pl}}\sin\left(\frac{m_\phi a}{k}x\right)\right] \notag\\
    &\times \p_x h_{R/L} + \frac{2a^2}{k^2M_{\rm pl}^2}\pi_{R/L}^T\,.
\end{align}
In Eq.~\eqref{FinalEOMh}, besides the matter part, the CS term also acquires a $k$-dependence. Since in the Boltzmann hierarchy, the collision term involves either a term $k/k_\star$ in the propagation of inflationary GWs, or $k\tau_i$ in the propagation of stochastic GWs, for numerical convenience in the analyses of the GWs spectra we can write, respectively, either $m_\phi a/k = (m_\phi a/k_\star)/(k/k_\star)$ or $(m_\phi a \tau_i)/(k\tau_i)$. 

The prefactor of the CS term, $(2\alpha H m_\phi) / M_{\rm pl}$, can be treated as a constant for specific cosmic time scales, e.g. either during or after reheating. During reheating, the inflaton oscillates and transfers energy to the matter fields, generating particles. For convenience, the inflaton mass $m_\phi$ can be regarded as approximately constant in this era. Recall the prefactor can be also recast as $(2\alpha \phi_0 m_\phi^2) / M_{\rm pl}^2$, where $\phi_0$ represents the inflaton amplitude at the end of slow-roll inflation, and thus it is effectively a constant.

After reheating, the parametric resonance ends, and the inflaton oscillation frequency satisfies $m_\phi < H$. Furthermore, in the radiation-dominated Universe, where the spacetime is nearly flat and $H \approx 0$, the inflaton mass $m_\phi$ also approaches zero. Consequently, the prefactor becomes negligible, allowing us to ignore the CS term after reheating.

Previous works (e.g., Refs.~\cite{Alexander:2004us,Alexander:2004wk,Alexander:2014bsa}) have already focused on GWs solutions in CS gravity, but without anisotropic stress-energy tensor. These studies show that the CS term introduces a damping factor $\sim e^{\pm\Theta x}$, which affects the amplitude of the right/left-handed gravitational waves. Notably, different values of $\Theta$ result in a constant difference in damping of the amplitude, without altering the shape of the GWs power spectrum. For numerical convenience, we set $|\Theta| = 10^{-2}$ and present the results for damping effect on the spectra of inflationary GWs re-entering the horizon during reheating, as well as the causal GWs generated during reheating.

During reheating, the inflaton transfers energy to the fermion fields. Given that neutrinos decoupled from weak interactions, and free-streamed at around $1$MeV, they are not expected to dampen the GWs either re-entering or generated during reheating. The standard model fraction of neutrinos $R_{\rm sm} = 0.40523$ cannot be exploited to select a value of $\Omega_\nu$. Other dark fermions must be taken into account to explain the origin of the damping effect. Nonetheless, we cannot choose any precise energy fraction of dark fermions that would be involved. To facilitate a direct comparison with the results in Ref.~\cite{Loverde:2022wih}, we then select $\Omega_\nu = 1/20$ as a reference value. The exact value of $\Omega_\nu$ can then be determined by the damped power spectrum if dark fermions are observed. For example, one can find that $\Omega_{\rm GW}/\Omega_{\rm GW}[f_{\rm fs}]\approx 0.64$, when $k/k_\star \ll 1$, for $\Omega_\nu = 0.40523$, as provided by neutrinos. To illustrate the quantitative role of the CS term, we fix the parameters by setting $n=5$, thus encompassing fermions being decoupled from the interactions. To inspect the case of re-coupling, we fix instead $n<2$. Then, after solving numerically the equations of motion, we are able to check that the CS term does still contribute to the power spectrum, and the results are similar than for $n=5$. 

\subsection{Inflationary gravitational waves}
Amplitudes of GWs formed during inflation are frozen at the time of horizon crossing. After the slow roll phase and during the Universe radiation dominated era, frozen amplitudes will re-enter the horizon and interact with particles. The initial conditions for the GWs can be summarized as it follows:
\begin{align}\label{initialin}
    \lim_{x\rightarrow 0} h_\lambda(x) = h_{\lambda,0},\quad \lim_{x\rightarrow 0}\p_xh_\lambda(x) = 0, \quad \lim_{x\rightarrow 0}F_{\lambda,l}=0\,.
\end{align}
\begin{figure*}[t]
    \centering
    \includegraphics[width=\textwidth]{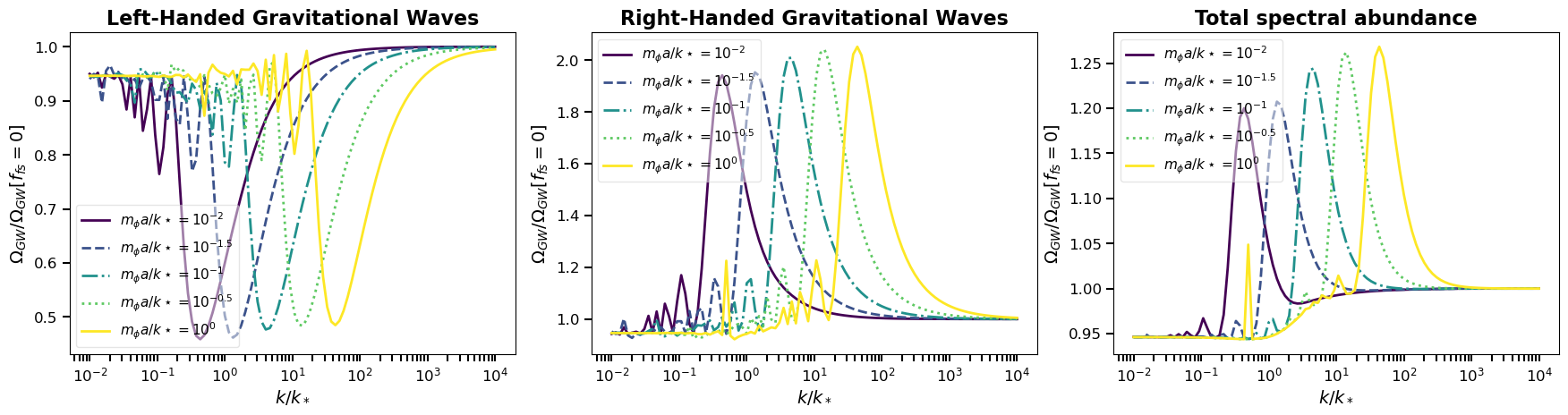}
    \caption{Plot of the power spectrum of GWs damped by (dark) fermions, with right or left helicities, for a species of dark fermions with energy fraction $\Omega_\nu = 1/20$. Within the images, the damping effect is described by the power spectrum $\Omega_{\rm GW}$, provided by the GWs interacting with fermions and normalized by the free GWs spectral abundance $\Omega_{\rm GW}[f_{\rm fs}]$, calculated at $f_{\rm fs}=0$, i.e. when the energy fraction of the free streaming fermions is zero. From the figure, the difference among the spectra for the right/left-handed GWs explicitly appears. Right/left-handed GWs show peaks/dips, whose position depends on the inflaton mass $m_\phi$ and is compared to the decoupling scale $k_\star$. The amplitudes of the peaks/dips depend on the prefactor $\Theta$ of the CS term. Peaks/dips appear because GWs mediate the parametric resonance of the particle production. In the right panel of the figure, we averaged over the left- and right-handed GWs power spectra, to show the total spectral abundance including damping effects.}
    \label{fig:dspectrumin}
\end{figure*}
In our analysis, the decoupling time of fermions, $\tau_\star = 1/\mathcal{H}_\star = 1/k_\star$, plays a crucial role in shaping the power spectrum when compared to the GWs re-entering time, $1/k$. We solve the equation of motion \eqref{FinalEOMh}, the Boltzmann hierarchy \eqref{BoltHier}, and the collision term \eqref{collisionterm} with the optical depth \eqref{InterRatek}, using the strategy outlined at the beginning of this section. In Fig.~\eqref{fig:dspectrumin}, we present the spectrum resulting from our analysis, and explore various choices for the inflaton mass in relation to the decoupling frequency $k_\star$. 

Our numerical results show that when the GWs frequency is much smaller than the decoupling frequency $k_\star$, i.e. when the corresponding modes re-enter the horizon after fermions have decoupled, the damping factor in the spectrum aligns with the findings of Ref.~\cite{Loverde:2022wih}. This indicates that at the time of horizon re-entry, fermions concerned are sufficiently decoupled and free-streaming, thereby damping GWs' amplitudes in agreement with the prediction provided by fully free-streaming fermions.\\

As the frequency increases, the overall features of the GWs spectra remain consistent with the results found by Ref.~\cite{Loverde:2022wih}, in which scenarios that only involve fermions, without CS term being added, were considered. However, an evident mismatch emerges between right- and left-handed GWs, along with a distinct oscillatory behavior in the spectrum as a function of frequency. To explain this oscillation, we refer to Ref.~\cite{Alexander:2014bsa}, which discusses gravitationally mediated parametric resonance. Following the arguments in Ref.~\cite{Alexander:2014bsa}, the phenomenon occurs when the oscillatory damping of the GWs' amplitude, induced by the CS term proportional to $\Theta$, becomes comparable to the oscillation rate of the inflaton. In this context, the inflaton transfers energy to the GWs, mediating the energy transfer to the matter field. This requires 
\begin{align} \label{condi}
    \frac{2\alpha \phi_0 m_\phi}{M_{\rm pl}^2} \sim \frac{m_\phi a}{k}\,,
\end{align}
as indicated by the dips and peaks in Fig.~\eqref{fig:dspectrumin}. Since $m_\phi a / k = (m_\phi a / k_\star) / (k / k_\star)$, if $2\alpha \phi_0 m_\phi / M_{\rm pl}^2$ remains constant, the previous condition is satisfied for larger values of $k$ when $(m_\phi a / k_\star)$ increases. Consequently, the dips and peaks appear later in the spectrum, consistently with the results shown in figure \eqref{fig:dspectrumin}. Thus, the larger is the mass of the inflaton, the higher is the frequency of the modes that dominates the particle production, and particles with higher energies are produced. The difference in the spectra of right- and left-handed GWs indicates that more energy is transferred to the right-handed GWs than is transferred from the latter to the matter field. In contrast, for the left-handed GWs, more energy is transferred to the matter field than is gained from the inflaton.
\begin{figure*}[t]
    \centering
    \includegraphics[width=\textwidth]{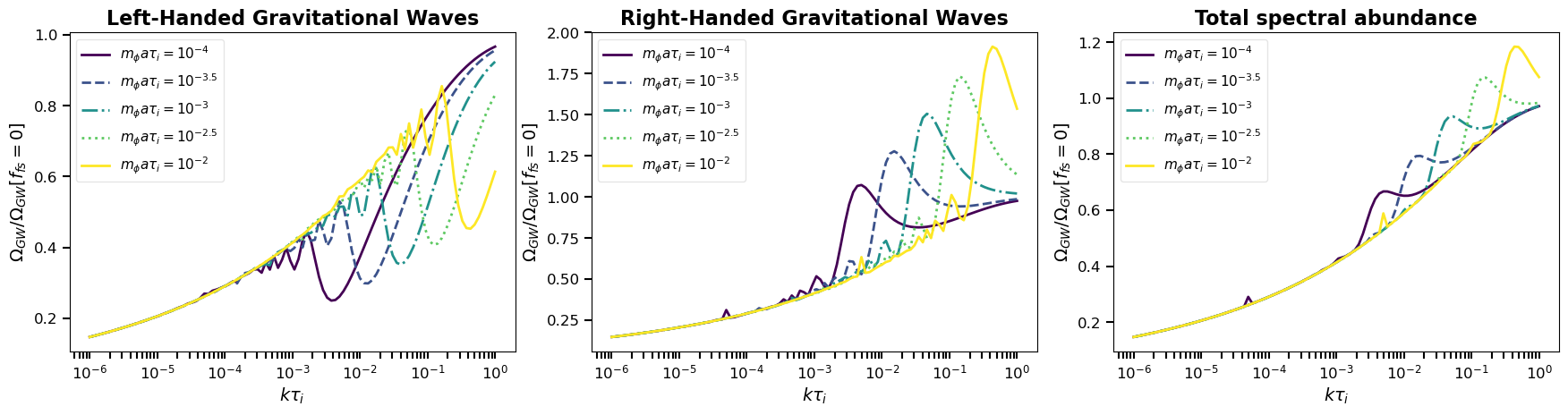}
    \caption{We plot the result of the power spectrum of right/left-handed causal GWs, in a similar way to Fig.~\eqref{fig:dspectrumin} - the damped left- and right-handed GWs power spectrum is plotted in the left and middle panels, respectively, while the total spectral abundance including damping effects is plotted in the right panel, averaged over both chiralities. The peaks and dips observed in the normalized power spectrum are generated in a similar manner to those in the inflationary GWs.}
    \label{fig:dspectrumre}
\end{figure*}

\subsection{Causal gravitational waves}
Causal GWs \cite{Hook:2020phx} are by definition influenced by causality, hence that their generation and propagation are constrained by the speed of light and produced by local physical processes, including phase transitions and parametric resonances in the early Universe. Causal GWs are generated by a sudden kick at a specific time $\tau_i$, and their initial amplitudes are zero. The initial conditions the read
\begin{align}\label{initialre}
    h_\lambda(\tau_i,\mathbf{k})=0, \quad \p_\tau h_\lambda(\tau_i, \mathbf{k}) = J_i, \quad F_{\lambda,l}(\tau_i,\mathbf{k}) = 0\,,
\end{align}
where $J_i$ is a source term. Even though causal GWs are produced inside the horizon, the wave numbers corresponding to low-frequency tails are super-horizon, and will be influenced by the free-streaming particles de/re-coupled from interactions. After the propagating modes are produced, the GWs' amplitudes are frozen to the constant value $h_\lambda \approx J_i/\mathcal{H}_i$ when the wave-number satisfies $k\tau_i \lesssim 10^{-3/2}$ \cite{Loverde:2022wih}. Then GWs solutions are normalized by a factor $J_i/\mathcal{H}_i = J_i\tau_i$. 

To solve for the GWs amplitude with these initial conditions, we have a few parameters to be fixed, namely the de-/recoupling times relative to the phase transition $\tau_\star/\tau_i = k_i/k_\star$, as indicated in Eq.~\eqref{InterRatei}. Since we are interested in the birefringence of GWs, we are not going to plot multiple figures with different values of $\tau_i/\tau_\star$, as depicted in Ref.~\cite{Loverde:2022wih}. For convenience, we set $\tau_i/\tau_\star=1$ and simplify the numerical analysis. The relevant quantities here are the mode frequencies, to be compared to the phase transition scale $1/\tau_i$. The resulting power spectrum is plotted with respect to $k\tau_i$, and shown in Fig.~\eqref{fig:dspectrumre}. We find that the shape of the spectrum shares a behavior similar to that shown in~\cite{Loverde:2022wih}, with $\Omega_\nu = 1/20$, and that the oscillations in the damped spectra are similar to the one studied in the inflationary caseRef.. Since we have set $k_i=k_\star$, we can compare the result with Fig.~\eqref{fig:dspectrumin}. If we look at the curves provided by $m_\phi a\tau_i = 10^{-2}$ (or $m_\phi a/k_\star = 10^{-2}$), we may distinguish dip/peak appearing around $k\tau_i=0.4$ (or $k/k_\star=0.4$). This clearly shows that the modes that dominate mediating parametric resonances do not depend on the GWs sources, as expected. However, since causal GWs are generated by local processes and are fixed by causality, they contain longer-wavelength modes than the inflationary GWs. These components finally contribute to the resonant production of particles characterized by longer wavelengths. For a fixed energy level $E$, relativistic particles produced by causal GWs and characterized by longer wavelengths are then forced to have larger values of their masses in order to fulfill the mass-shell condition. This is a different case from inflationary GWs.

\subsection{Evolution of the gravitational waves}
\begin{figure}[t]
    \centering
    \includegraphics[width=0.45\textwidth]{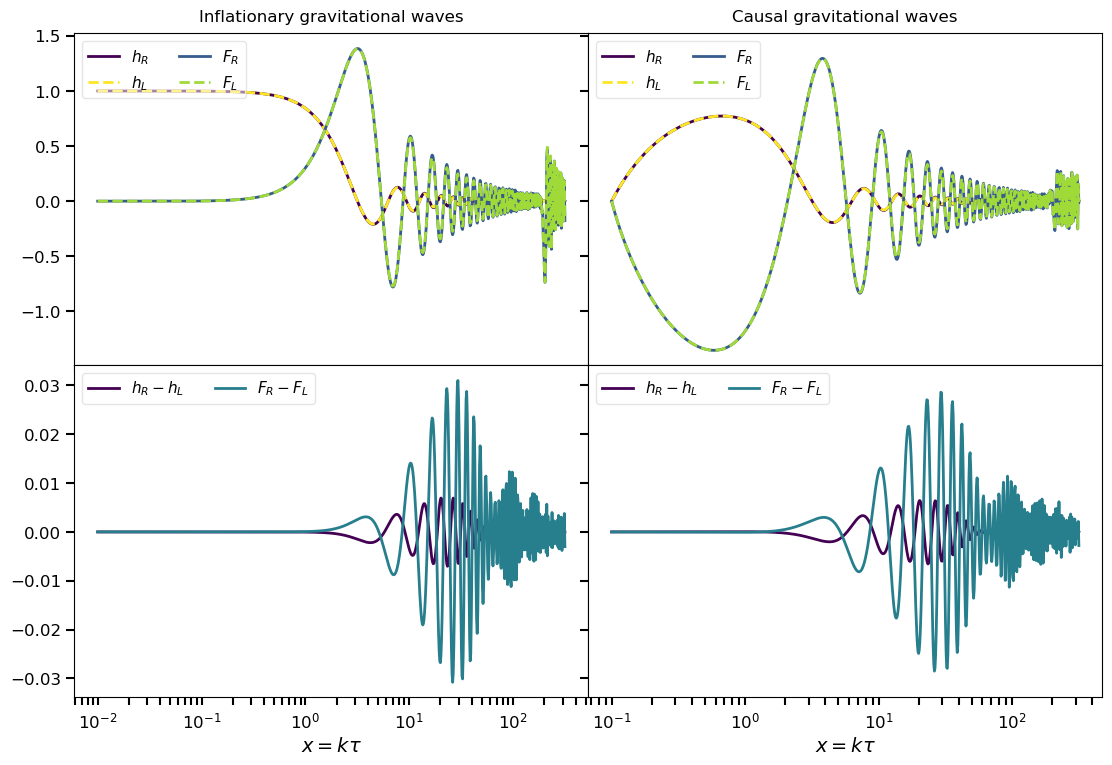}
    \caption{GWs' evolution and zeroth component of the Boltzmann hierarchy for different chiralities. The two figures on the top separately depict the evolutions of the amplitudes, the solid line denoting the right-handed components and the dashed line the left-handed components. Curves with different chiralities are overlapped and it is not possible to distinguish them. On the bottom, we plotted the difference between the right/left-handed components. Now the difference between right- and left-handed components becomes manifest, with a typical time oscillation after re-entering the horizon.}
    \label{fig:evolutions}
\end{figure}    

Before drawing our conclusion, we elaborate here on the time evolution of the GWs amplitude and phase-space distribution function. Solving the equations of motion, with the initial conditions \eqref{initialin} and \eqref{initialre}, we plot the evolution of $h_R$, $h_L$, $F_{0,R}$ and $F_{0,L}$ with a specific set of parameters, provided by $k/k_\star=k\tau_i = 10^{-2}$, the other parameters acquiring the same values formerly fixed. The result is shown in Fig.~\eqref{fig:evolutions}. In the upper plots, chiral components are shown separately: the right-handed (solid curves) GWs' components and the distribution functions' zeroth component almost overlap the left-handed (dashed curves) GWs' components. Conversely, in the lower plots the difference between the right- and left-handed curves is plotted, and the difference among the chiral components becomes visible, featuring a typical oscillation when the horizon re-enters. Combining this observation with the results in Fig.~\eqref{fig:dspectrumin} and Fig.~\eqref{fig:dspectrumre}, we may claim that birefringence exists in both the GWs and the phase-space distribution function for the matter field. Indeed, even though it may be hard to be distinguish directly their time evolutions for small values of the CS pre-factor $\Theta$ --- see e.g. the case in which $\Theta\sim 10^{-2}$ --- the damping effect provided by different chiralities on the power spectra is explicit\footnote{We considered also larger values of $\Theta$, including $\Theta\sim 1$. As expected, the time evolution of the right-handed GWs components and the distribution functions can also be explicitly distinguished also in this case from the left-handed GWs components.}.  As a result, stochastic GW power spectra characterized by a damping effect owing to relativistic fermions provide a reliable probe for testing CS gravity.

\section{Discussions} \label{conclu}
\noindent 
We have discussed the damping effect provided by self-interacting relativistic fermions in Chern-Simons gravity, where a birefringence in the gravitational waves is introduced. The damping effect induces a reshape of the power spectrum of the gravitational waves. The Chern-Simons term significantly affects the power spectrum during reheating, even though the difference among the right/left-handed gravitational waves amplitudes is very small. Thus, the difference in the chiral components of the gravitational waves power spectra provides an optimal probe to test Chern-Simons gravity. In light of the results shown in Fig.~\eqref{fig:dspectrumin} and in Fig.~\eqref{fig:dspectrumre}, we may search for similar effects in the power spectra of the stochastic gravitational waves signals from the early Universe. The parameter space corresponding to the Chern-Simons coupling $\alpha$ and the inflaton mass $m_\phi$ can be scanned by identifying the amplitudes and the positions of the dips/peaks on the power spectra, individuated by $2\alpha H m_\phi/M_{\rm pl}$ and $m_\phi a/k$, respectively. This also provides a useful phenomenological scheme to study the physics of inflation. When the value of $m_\phi$ is constrained, using the Friedmann equation we may constrain the end time of inflation. Furthermore, the Chern-Simons term results in the gravitational waves mediated parametric resonant particle production. The related reshaping of the power spectra finally provides a direct evidence for reheating, with an effect of parametric resonance.\\

A specific application of our results hinges on the realization that effective Chern-Simons gravity can be derived from self-interacting fermions\footnote{In Refs.~\cite{Alexander:2014eva,Bambi:2014uua,Alexander:2014uaa,Addazi:2016rnz,Addazi:2017qus,Addazi:2018zjv}, the cosmological and astrophysical implications of an effective model involving self-interacting fermions were studied. This can be achieved by integrating out in the path integral the torsional component of the Lorentz connection in the Einstein-Cartan-Holst action coupled to Dirac fermions.}, as discussed in Ref.~\cite{Alexander:2022cow}. Fermions that are subject to beyond-standard model interactions can be provided by dark ones, as discussed in Refs.~\cite{PhysRevD.83.115009,Banerjee:2017wxi,Banerjee:2021hfo,Ghosh_2022,PhysRevD.106.063013}. In general, the self-interacting terms involve the bilinears 
\begin{align}
(\bar{\psi}\psi)^2,\ (i\bar{\psi}\gamma^5\psi)^2,\ (\bar{\psi}\gamma^\mu\psi)^2,\ (\bar{\psi}\gamma^\mu\gamma^5\psi)^2,\ (\bar{\psi}\sigma^{\mu\nu}\psi)^2,
\end{align}
where $\gamma^I$ are the gamma matrices, and $\sigma^{\mu\nu} = [\gamma^\mu,\gamma^\nu]$. Using the Fierz identities and the Pauli-Kofink relation \cite{PhysRev.140.B1467}, one can find that the only independent terms are the scalar $(\bar{\psi}\psi)^2$ and the pseudoscalar $(i\bar{\psi}\gamma^5\psi)^2$ ones. \\

In Ref.~\cite{Alexander:2022cow}, the emergence of Chern-Simons gravity is a consequence of the scalar and pseudoscalar terms, in which the decay constant $f=1/\alpha$ is determined by the mass of fermions. As a result, the damping effect provided by the self-interacting fermions, when considered together with birefringence in the gravitational waves, provides a way to probe dark matter. To find the role of the scalar and pseudoscalar interaction terms for a sizable damping effect in gravitational waves, we need to model the collision term \eqref{collisionterm} more specifically.  We leave this task to a future investigation, and observe that if the birefringence was detected in the power spectrum, we could derive a constraint for the masses of the dark fermions or the (sterile) neutrinos that are self-interacting. \\

Finally, the difference between the right- and left-handed gravitational wave power spectra implies that right-handed gravitational waves store more energy than left-handed gravitational waves during the mediation of the parametric resonance. Consequently, we expect that it will be possible to relate this feature to the chiral asymmetry in the particle sector, e.g. neutrinos. This is a complicated theoretical analysis that we are pursuing in separate ongoing projects.\\

The possibility of detecting the signals of birefringence in stochastic gravitational wave power spectra is of significant importance. A promising approach to detect inflationary or causal gravitational waves is based on the Pulsar Timing Array (PTA). This probe plays a crucial role in the B-mode polarization of the Cosmic Microwave Background (CMB). As discussed in Ref.~\cite{Loverde:2022wih}, the decoupling or recoupling of self-interacting neutrinos leaves a substantial impact on the B-mode angular power spectrum in the CMB, which subsequently influences the signals of gravitational waves. For the gravitational wave power spectrum, Ref.~\cite{Loverde:2022wih} also explores the signals that can be observed today for inflationary gravitational waves delayed by the decoupling of self-interacting neutrinos, in conjunction with observational results obtained from several programs. \\

The PTA and interferometers serve as a high-frequency lever arm for measuring the inflationary or causal gravitational wave background, thereby providing evidence for the decoupling of new particles from interactions. In addition, astrometry, which involves precise measurements of star motions, offers an alternative avenue for cross-checking the PTA results and investigating low-frequency ($\ll$ nHz) gravitational waves and their chirality through the spatial deflection of photons. This approach serves as a complementary method to the PTA, as exemplified, e.g., in Refs.~\cite{Moore:2017ity,Klioner:2017asb,Garcia-Bellido:2021zgu,Caliskan:2023cqm}. The Gaia mission \cite{gaia2016gaia}, along with the upcoming missions Roman \cite{Wang:2020pmf,Wang:2022sxn,Haiman:2023drc,Pardo:2023cag} and Theia \cite{boehm2017theia,malbet2021faint}, provides astrometric observations.\\

To provide a more specific example, the deflection angle of distant sources in a particular direction labeled $i$, denoted as $\delta n^i$, can be calculated using the astrometric effect caused by the stochastic gravitational wave background. This calculation is detailed in Ref.~\cite{Book:2010pf} employing tetrads. Tetrads enable the coupling of self-interacting Dirac fermions to gravity, resulting in birefringence in gravitational waves. Consequently, the deflection angles will differ when the astrometric effect is attributed to gravitational waves with different chiralities. The correlation function of $\delta n^i$ is linearly related to the power spectrum of gravitational waves that can be observed today \cite{Book:2010pf}. In Ref.~\cite{Caliskan:2023cqm}, the method for detecting the power spectrum of gravitational waves with different polarizations is discussed, which yields correlations between two signals received from a pair of pulsars, stars, or pulsar-star pairs, leading to the well-known Hellings-Downs curve in PTA with redshift correlations. However, this method remains a potential approach, and extensive research is necessary in both the theoretical and observational domains to further develop and refine it.\\

We finally end this discussion by emphasizing that the peaks and dips predicted within this model, being of the same order of the power spectrum, are in principle observable by next generation of space-borne gravitational interferometers, including LISA, Taiji and Tianqin. Indeed, even though these experiments are not intended to distinguish the chiral components of the gravitational radiation, nonetheless the sum of the gravitational powers spectra characterized by different chirality will be still featuring observable peaks and dips.

\section*{Acknowledgements}
The authors wish to thank Tucker Manton for discussions and comments. This work utilized the Python packages \textsf{NumPy} \cite{Harris:2020xlr}, \textsf{SciPy} \cite{Virtanen:2019joe}, \textsf{matplotlib} \cite{Hunter:2007ouj}, and \textsf{Joblib} \cite{joblib}. SA acknowledges support by the Simons Foundation through Award No. 896696. JL and AM acknowledges the support by the NSFC, through the grant No.\ 11875113, the Shanghai Municipality, through the grant No.\ KBH1512299, and by Fudan University, through the grant No.\ JJH1512105.

\bibliographystyle{elsarticle-num}
\bibliography{references}

\end{document}